\documentclass[preprint,showpacs,preprintnumbers,amsmath,nofootinbib]{revtex4}
\usepackage{amsfonts}
\usepackage{amssymb}
\usepackage{latexsym}
\usepackage{fancybox}
\usepackage{graphicx}
\begin{document}

\preprint{M\'exico ICN-UNAM, \, September 2007}

\title{Hydrogen atom in a  magnetic field: electromagnetic transitions of the lowest states}

\author{J.C. \surname{L\'opez Vieyra}}
\email{vieyra@nucleares.unam.mx}
\affiliation{}

\author{H.~O.~Pil\'on}
\email{horop@nucleares.unam.mx}

\affiliation{Instituto de Ciencias Nucleares, UNAM, Apartado
Postal 70-543, 04510 M\'exico}

\date{September 21, 2007}

\begin{abstract}
A detailed study of the lowest states $1s_0, 2p_{-1} ,2p_0$ of the hydrogen atom placed in a magnetic field $B\in(0-4.414\times 10^{13}\,{\rm G})$  and their  electromagnetic transitions ($1s_{0}   \leftrightarrow 2p_{-1}$ and  $ 1s_{0}  \leftrightarrow 2p_{0}$)
is carried out in the Born Oppenheimer approximation. The variational method is used with a physically motivated  recipe to design simple trial functions applicable to the whole domain of magnetic fields. We show that the proposed functions yield very accurate results for the ionization (binding) energies. Dipole and oscillator strengths are  in good agreement with results by Ruder {\em et al.}~\cite{Ruderbook} although we observe deviations up to $\sim 30\%$   for the oscillator strength of the (linearly polarized)  electromagnetic transition $1s_{0}   \leftrightarrow 2p_{0}$  at strong magnetic fields $B\gtrsim 1000$\,a.u.
 \end{abstract}


\maketitle

\section{Introduction}
\label{sec:intro}

Contemporary  X-ray space observatories, like Chandra, XMM-Newton and their predecessors, have  collected a considerable amount of observational data of the thermal emission  coming from surface layers of neutron stars which are characterized by enormous  magnetic fields  $B\sim 10^{12}-10^{13}\,$G  (see \cite{Pavlov2001},\cite{Zavlin:2007}). In particular, the observation of  absorption features in the X-ray spectrum of some isolated neutron stars (see {\em e.g.} ~\cite{Sanwal:2002,Kerkwijk:2004}) has suggested  possible models of atmospheres which allow the presence of Coulomb systems ~\cite{Turbiner:2004m, Turbiner:2005m,Mori:2007,Heyl:2007}.
 The hydrogen atom is the simplest and most studied   Coulombic system in weak
and strong magnetic fields (see for example the early review~\cite{Garstang} and references therein, and \cite{Ruderbook,K&L:ET,Karnakov:2003} for more recent studies).

In the present study our goal is to apply a physics recipe (described in full
generality in~\cite{Tur}) for choosing variational trial functions to the
study the hydrogen atom in a magnetic field and its electromagnetic
transitions between the lowest bound states $1s_{0}$, $2p_{-1}$ and $2p_{0}$.
The study is intended as a test of the methodology developed in~\cite{Tur}.
Electromagnetic transitions in the hydrogen atom in the absence of a magnetic
field is a widely described subject (see {\em e.g.} \cite{Sobelman}). In a
strong magnetic field such electromagnetic transitions have been studied by a
number of authors, specially focusing on the effects of the transverse motion
across the magnetic field direction (see {\em e.g.}
\cite{CuvelliezBaye,PP95,Potekhin94,Potekhin98}).

Our consideration is non-relativistic, based on a variational solution of the Schroedinger equation. Thus, the magnetic field strength is
restricted by the Schwinger limit $B=4.414\times 10^{13}$\,G.
Our study is also based on the Born-Oppenheimer approximation of zero order:
the proton  is assumed to be infinitely massive. Thus we neglect the effects of the CM motion {\it i.e.}  the effects of the transverse motion of the atom with respect to the magnetic field orientation.

The study is realized in two steps: (i) a variational calculation of the states $1s_{0}$, $2p_{-1}$ and  $2p_{0}$  is done with suitable trial functions (selected according to the physics recipe), and (ii) with the variationally obtained approximate wavefunctions  we calculate the allowed radiative transitions among these states in the electric dipole approximation (see below).

Atomic units are used throughout ($\hbar$=$m_e$=$e$=1), albeit
energies are expressed in Rydberg (Ry). The magnetic field $B$ is
given in a.u. with $B_0= 2.35 \times 10^9$\,G, although frequently we will also use  magnetic fields strengths given
in Gauss for convenience.

\subsection{Hamiltonian}
\label{sec:Hamiltonian}
The Hamiltonian which describes the Coulomb system formed by an infinitely massive proton and one electron $(pe)$  placed in a homogeneous constant magnetic field directed along the $z$-axis, $\nobreak{{\bf B}=(0,0,B)}$ is given by\
\begin{equation}
\label{Ham} \hat{\mathcal {H}}  = ({\hat {\mathbf p}}+{\cal A})^2  -\frac{2}{r} \ = \
\hat{\mathbf p}^2  -\frac{2}{r} + (\hat{\mathbf p} \cdot {\cal A}+{\cal A}\cdot\hat{\mathbf p}  ) +
{\cal A}^2 \ ,
\end{equation}
where $\hat{\mathbf p}  =-i \nabla$ is the electron momentum,  $r$ is the distance between the electron and the proton fixed at the origin, and ${\cal A}$ is a vector potential  corresponding to the magnetic field $\bf B$. A contribution to the energy coming from the coupling between the electron intrinsic magnetic moment and the magnetic field $\sim {\hat {\mathbf S}}\cdot {\mathbf B}$, being  constant, has been  dropped from (\ref{Ham}). Now, if we choose the vector potential in the  symmetric gauge
\[
   {\cal A}= \frac{B}{2} (-y,\ x,\ 0)\ ,
\]
the Hamiltonian acquires the form
\begin{equation}
 \label{ham-H}
         \hat{\mathcal {H}}= -\mathbf{\Delta} - \frac{2}{r} + {\hat{l}}_{z}B +
         \frac{B^2\rho^2}{4}\ ,
\end{equation}
where $\mathbf{\Delta}$ is the Laplacian operator, and ${\hat{l}}_{z}$ is the conserved $z$-component of the electron angular momentum. The Hamiltonian (\ref{ham-H}) is also invariant with respect to  the $z$-parity $\pi_z$   ({\em i.e.} reflections $z\to -z$). Thus the eigenstates can be classified by $(m,\pi_z)$: the magnetic quantum number $m$, corresponding to the conservation of ${\hat{l}}_{z}$, and the $z$-parity quantum number $\pi_z=\pm 1$.

\subsection{Choice of trial functions}
\label{sec:trialfchoice}

The procedure which we use to explore the problem, is the variational  method  with a well defined recipe for choosing trial functions. This recipe is based on physical arguments,
described in full generality in \cite{Tur}. The basic ingredients are (i) for a given trial function $\psi_{trial}$, the potential $V_{trial}=\mathbf{\Delta}\psi_{trial}/\psi_{trial}$, for which such function is an exact eigenfunction, should reproduce as many as possible the basic properties of the original potential, {\em e.g.} in the present case it should reproduce the Coulomb singularities and the harmonic oscillator behavior at small and large $r$ distances  respectively, and (ii)  the trial function $\psi_{trial}$ should include the symmetries of the problem. For example, if the ground state is studied, the trial function $\psi_{trial}$ has to be a nodeless function.  Adhering to this recipe, in~\cite{PotekhinTurbiner} it was proposed  the following function for the hydrogen ground state $1s_0$
  \begin{eqnarray}
\label{trialgs}
  \lefteqn{ \Psi_{1s_{0}}=} \\[5pt]
 &&
 e^{-\sqrt {\gamma_{1} r^{2} + (\gamma_{2}r^{3} + \gamma_{3}\rho^{2}r+
\gamma_{4}\rho^{3}+\gamma_{5}\rho r^{2})B^2  + \frac{\gamma_{6}B^{2}\rho^{4}}{16}
+ \frac{\gamma_{7}B^{2}\rho^{2}r^{2}}{16} }} \nonumber\,,
  \end{eqnarray}
where $\gamma_1,\ldots \gamma_7$ are variational parameters. The potential associated with the function (\ref{trialgs}) reproduces the Coulomb singularity at $r\to 0$ as well as the harmonic oscillator for $r\to \infty$ of the original Hamiltonian ({\ref{ham-H}}). The function (\ref{trialgs}) has no nodes. In the following we will use the function (\ref{trialgs}) for the variational study of the ground state $1s_{0}$.

For the lowest excited states $2p_{-1}$, $2p_{0}$ the presence of the magnetic field does not modify the nodal structure of the field-free exact eigenfunctions\footnote{{\em i.e.} the nodal surfaces defined by the condition $\psi_{trial}(r,\theta,\phi)=0$.}. A hint for this can be obtained in first order degenerate perturbation theory in $B^2$, where one can see that the states $2p_{-1}$, $2p_{0}$
are not mixed. Thus   the following variational functions for the  $2p_{-1}$, $2p_{0}$ states are proposed

 \begin{subequations}
\begin{eqnarray}
\label{trial2pm1}
 \Psi_{2p_{-1}}&=& \rho e^{-i\,\phi}\,\psi_0 \,, \\
  \label{trial2p0}
\Psi_{2p_{0}}&=& z\,\psi_0 \,,
\end{eqnarray}
\end{subequations}
 where the functions $\psi_0$ have the same functional form as (\ref{trialgs}) but with their own $\gamma$-parameters.
 Thus, in the  functions (\ref{trial2pm1}), (\ref{trial2p0})  we keep the same polynomial
 prefactor  as in the corresponding field-free wavefunctions and multiply them by  a nodeless function. It is easy to check that the functions  (\ref{trialgs}), (\ref{trial2pm1}) and (\ref{trial2p0})  are orthogonal. These functions
describe the lowest energy states, $1s_0, 2p_{-1}, 2p_{0}$, among the states with quantum numbers $(m,\pi_z)=(0,+),(-1,+),(0,-)$ respectively, in all the range of magnetic fields studied.

\section{Electromagnetic transitions}
\label{sec:trans}

In a magnetic field the $m$-degeneracy of the hydrogen energy levels is fully removed and  electromagnetic transitions depend explicitly on the magnetic quantum numbers  of the initial and final states in the transition. A consideration of electromagnetic transitions in the electric dipole approximation is valid even in the case of high magnetic fields as long as transitions occur among states with the same Landau quantum number  (in this case the characteristic wave lengths are always much larger than the (longitudinal or transverse) {\em size}  of the system (see \cite{Ruderbook})). All the states   considered in the present study belong to the same ground Landau-level and  thus  the electric dipole approximation is justified. Relevant formulas for the electromagnetic transitions in the hydrogen atom in a magnetic field were given in \cite{Ruderpaper}. In the electric dipole approximation we are interested in the square of the matrix element (called dipole strength)
  \begin{equation}
\label{dtautau}
    d_{\tau',\tau}^{(q)}=|p_{\tau',\tau}^{(q)}|^{2} = |\langle
    \tau'|r^{(q)}|\tau \rangle|^{2}\,,
  \end{equation}
where $\tau',\tau$ label the final and initial states in the transition, and
$r^{(q)}$ are the spherical components ($q = 0, \pm 1$) of the electric dipole operator.
The oscillator strength of the transition is given by
   \begin{equation}
\label{ftautau}
    f_{\tau',\tau}^{(q)} = (E^{\tau'}_{b} - E^{\tau}_{b})
    |p_{\tau',\tau}^{(q)}|^{2}\,,
   \end{equation}
where $(E^{\tau'}_{b} - E^{\tau}_{b})$ is the (binding) energy difference of the initial and final states.
The transition probability is calculated according to the relation
   \begin{equation}
\label{wtautau}
    w_{\tau',\tau}^{(q)} = \frac{1}{3\tau_0} (E^{\tau'}_{b} -
       E^{\tau}_{b})^{3} |p_{\tau',\tau}^{(q)}|^{2}\,,
   \end{equation}
where $\frac{1}{\tau_0}=8\textrm{.}03\times 10^{9} \sec^{-1} $. We have two selection rules implicit in the matrix element $p_{\tau',\tau}^{(q)}$ (eq.~(\ref{dtautau})) {\em viz.}, parity change and
  \begin{equation}
    \Delta m = 0, \pm 1\,,
  \end{equation}
which implies $q=\Delta m$. Thus, the transitions with  $\Delta m =0$   are  characterized by  a linearly polarized radiation along the magnetic field direction with $q=0$, while  the transitions with  $\Delta m =1$   are characterized by  circularly polarized radiation with $q=+1$ (for right polarization), or $q=-1$ (for left polarization).

\section{Results}

\subsection{Binding Energies}
\begin{figure}[ht]
\begin{center}
 \includegraphics[angle=-90,width=300pt]{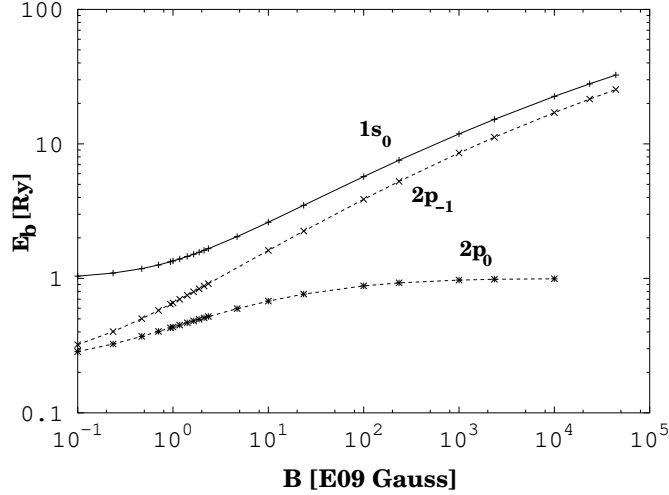}
\end{center}
\caption{\label{fig:1}Binding energies $E_b$ for the three lower states $1s_0, 2p_{-1},2p_0$  of the hydrogen atom as functions of magnetic field strength $B$. The curves show the calculated values of $E_b$ using the variational function (\ref{trialgs}) for the ground state $1s_0$ (marked by $+$), the function (\ref{trial2pm1}) for the state $1p_{-1}$ (marked by $\times$), and the function (\ref{trial2p0}) for the state $1p_0$ (marked by $\ast$). The points orresponding to the same state are joined by line-segments.}
\end{figure}

The results of the variational calculations of the total $(E_T)$ and binding\footnote{The binding energy is defined as the energy difference between the energy of a free electron in the magnetic field $B$ and the total energy $E_T$, {\em i.e.} $E_b=B(1+|m|+m)-E_T$ (for states in the ground Landau-level).} $(E_b)$
energies  for the lowest $1s_0$, $2p_{-1}$  and $2p_0$ states of the hydrogen atom in a  magnetic field ranging $ 0.235 \times 10^{9}\,{\rm G} \le B \le 4.414 \times 10^{13}$\,G  are presented in Tables~(\ref{Tabla:1S0}), (\ref{Tabla:2P-1}) and (\ref{Tabla:2P0}) respectively\footnote{The case of the ground state $1s_0$ was analyzed in \cite{PotekhinTurbiner} with the trial function (\ref{trialgs}). However, in order to have precise numerical information of the variational parameters appearing in the trial function, we repeated the calculations done in \cite{PotekhinTurbiner} for all the magnetic fields quoted there.}. Results for binding energies of these lowest states are also summarized in Fig.~(\ref{fig:1}).  In this figure one can immediately see that the binding energy grows steadily as the magnetic field increases for all three states $1s_0$, $2p_{-1}$ and $2p_0$. In particular, the ionization energy reaches $\sim 0.4$keV for a magnetic field at the Schwinger limit where $B=4.414 \times 10^{13}$\,G. The increasing in binding energy as the magnetic field grows is {\em faster}  (and comparable) for the  bound states $1s_0$, and $2p_{-1}$ (increasing in $\sim 20$ times for the domain $B\sim 10^9 - 10^{13}$\,G) as compared with the rate of increase of the binding energy for the bound state $2p_0$ (increasing only in $\sim 2$ times for the domain $B\sim 10^9 - 10^{13}$\,G). In fact the binding energy of the state  $2p_0$    approaches asymptotically to the value $E_b=1\,$Ry as $B\to\infty$ (see~\cite{Rau:1976} and Table~(\ref{Tabla:2P0})).

 For all states studied the results of the binding energies given by the simple trial
 functions (\ref{trialgs}), (\ref{trial2pm1}), (\ref{trial2p0})  are, in general, in very good agreement with the   approaches of Ruder {\it et al.}~\cite{Ruderbook} and
 Kravchenko {\it et al.}~\cite{K&L:ET}. For small to moderately high magnetic fields
 ($B\lesssim 1$\,a.u.)  the  relative differences between our binding energies and those
 of \cite{Ruderbook,K&L:ET} are found to be  $\lesssim 10^{-4}$.  It is worth to
 emphasize the remarkable coincidence in 9   digits for the ground state binding energy
 given by (\ref{trialgs}) and the today's most accurate results of  \cite{K&L:ET} in the
 domain of magnetic fields $B\lesssim 0.1$\,a.u.
The agreement of the binding energies with the corresponding perturbative results  \( E^{1s_0}_{b} \simeq 1+B-\frac{1}{2}B^{2}, \,\, \,\,E^{2p_{-1}}_{b}\simeq  \frac{1}{4}+2B-6B^{2},\,\,
E^{2p_0}_{b}    \simeq  \frac{1}{4}+B-3B^{2} \,\,\), obtained with a logarithmic perturbation theory (see \cite{Tur,Turbiner:1991,Horaciothesis} and references therein), is also very good.

However, the results for the binding energies of the three states studied
   obtained with the variational functions (\ref{trialgs}), (\ref{trial2pm1}) and
 (\ref{trial2p0}) show that the accuracy
gradually decreases as the magnetic field increases and the relative differences for such binding energies (when compared with the corresponding results of references \cite{Ruderbook} and \cite{K&L:ET}) reach  values $\sim 10^{-2}$ for $B\gtrsim 1000\,$a.u. (see Tables~(\ref{Tabla:1S0}), (\ref{Tabla:2P-1}), (\ref{Tabla:2P0})). It  indicates that the adiabatic separation of the transverse and longitudinal degrees of freedom is slightly delayed in the functions (\ref{trialgs}), (\ref{trial2pm1}) and
 (\ref{trial2p0}).

A comparison of the binding energies of the two lowest states $1s_0$ ($m=0$) and $2p_{-1}$ ($m=-1$) with the asymptotic (adiabatic) formulas $E_b^{(m)}\simeq \log^2 \frac{B}{\sqrt{2|m|+1}}$ (see {\em e.g.}~\cite{K&L:ET}) shows that, in both cases, the results given by this asymptotic formula are still far  from  the more accurate variational calculations, being different in about 3 times for the highest magnetic fields studied. For instance, at the magnetic field $B=10000$\,a.u.  the adiabatic formula gives $E_b^{1s_0}\simeq 85$\,Ry and $E_b^{2p_{-1}}\simeq 75$\,Ry, while the present numerical results are $E_b^{1s_{0}}= 27.96$\,Ry  and $E_b^{2p_{-1}}= 21.56$\,Ry, respectively (see Tables~(\ref{Tabla:1S0}),(\ref{Tabla:2P-1})). Even the asymptotic binding energy difference $\Delta E_b^{\rm asympt}\simeq 9.8$\,Ry is about 1.5 times larger than the variational one $\Delta E_b^{\rm var}\simeq 6.4$\,Ry for such a magnetic field.

In practice, a full variational  calculation is easily  done in a standard desktop computer,  it takes  few minutes of CPU time.

\begin{figure}[ht]
\begin{center}
 \includegraphics[angle=-90,width=300pt]{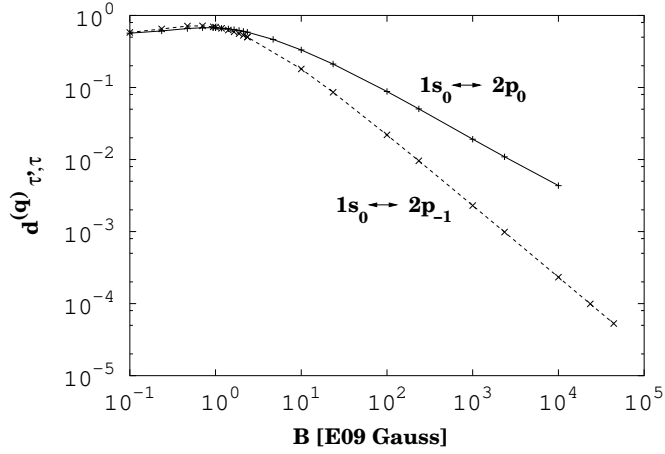}
       \caption{\label{fig:DS} Dipole strengths $d^{(q)}_{\tau'\tau}$ for the transitions
$1s_{0}  \leftrightarrow 2p_{-1}   $, and
$1s_{0}  \leftrightarrow  2p_{0}   $ as functions of magnetic field strength $B$. The curves show the calculated values of the dipole strengths $d^{(+1)}_{1s_{0}   \leftrightarrow 2p_{-1}}$ (marked by $\times$) and  $d^{(0)}_{1s_{0}   \leftrightarrow 2p_{0}}$ (marked by $+$) joined by line-segments.}.
\end{center}
\end{figure}

\begin{figure}[ht]
\begin{center}
 \includegraphics[angle=-90,width=300pt]{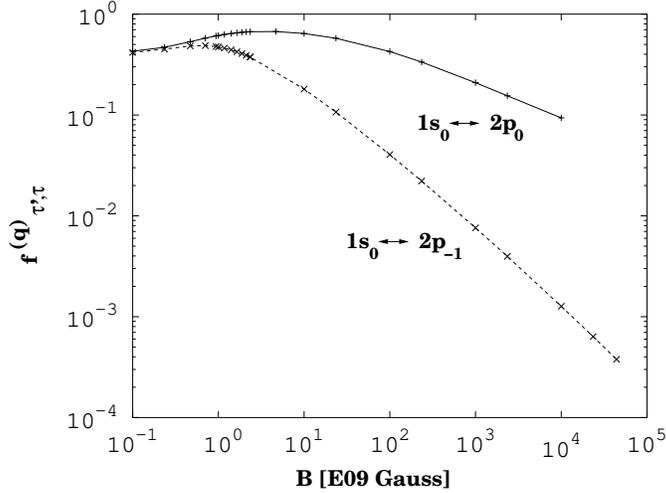}
\end{center}
 \caption{\label{fig:OS} Oscillator strengths $f^{(q)}_{\tau'\tau}$ for the transitions
$1s_{0}  \leftrightarrow 2p_{-1}   $, and
$1s_{0}  \leftrightarrow  2p_{0}   $ as functions of magnetic field strength $B$. The curves show the calculated values of the oscillator strengths $f^{(+1)}_{1s_{0}   \leftrightarrow 2p_{-1}}$ (marked with the symbol $\times$) and  $f^{(0)}_{1s_{0}   \leftrightarrow 2p_{0}}$ (marked with the symbol $+$)  joined by line-segments.}
\end{figure}

\begin{figure}[ht]
\begin{center}
\includegraphics[angle=-90,width=300pt]{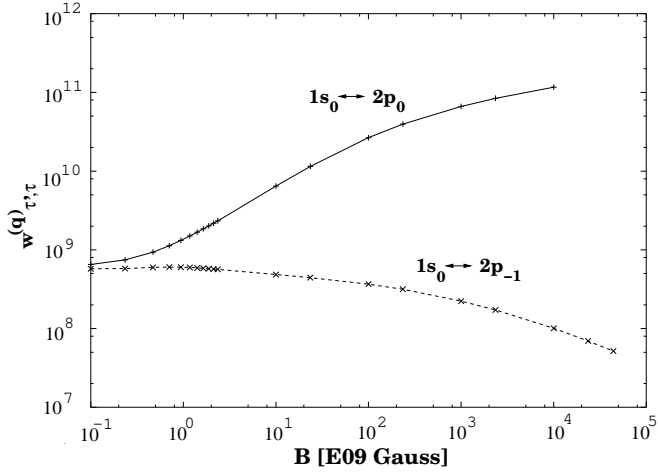}
\end{center}
 \caption{\label{fig:w} Transition probabilities $w^{(q)}_{\tau'\tau}$ for
the transitions $1s_{0}  \leftrightarrow 2p_{-1}   $, and
$1s_{0}  \leftrightarrow  2p_{0}  $. The curves show the calculated values of  the transition probabilities $w^{(+1)}_{1s_{0}   \leftrightarrow 2p_{-1}}$ (marked with the symbol $\times$) and  $w^{(0)}_{1s_{0}   \leftrightarrow 2p_{0}}$ (marked with the symbol $+$)  joined by line-segments.}
\end{figure}

\begin{figure}[ht]
\begin{center}
 \includegraphics[angle=-90,width=300pt]{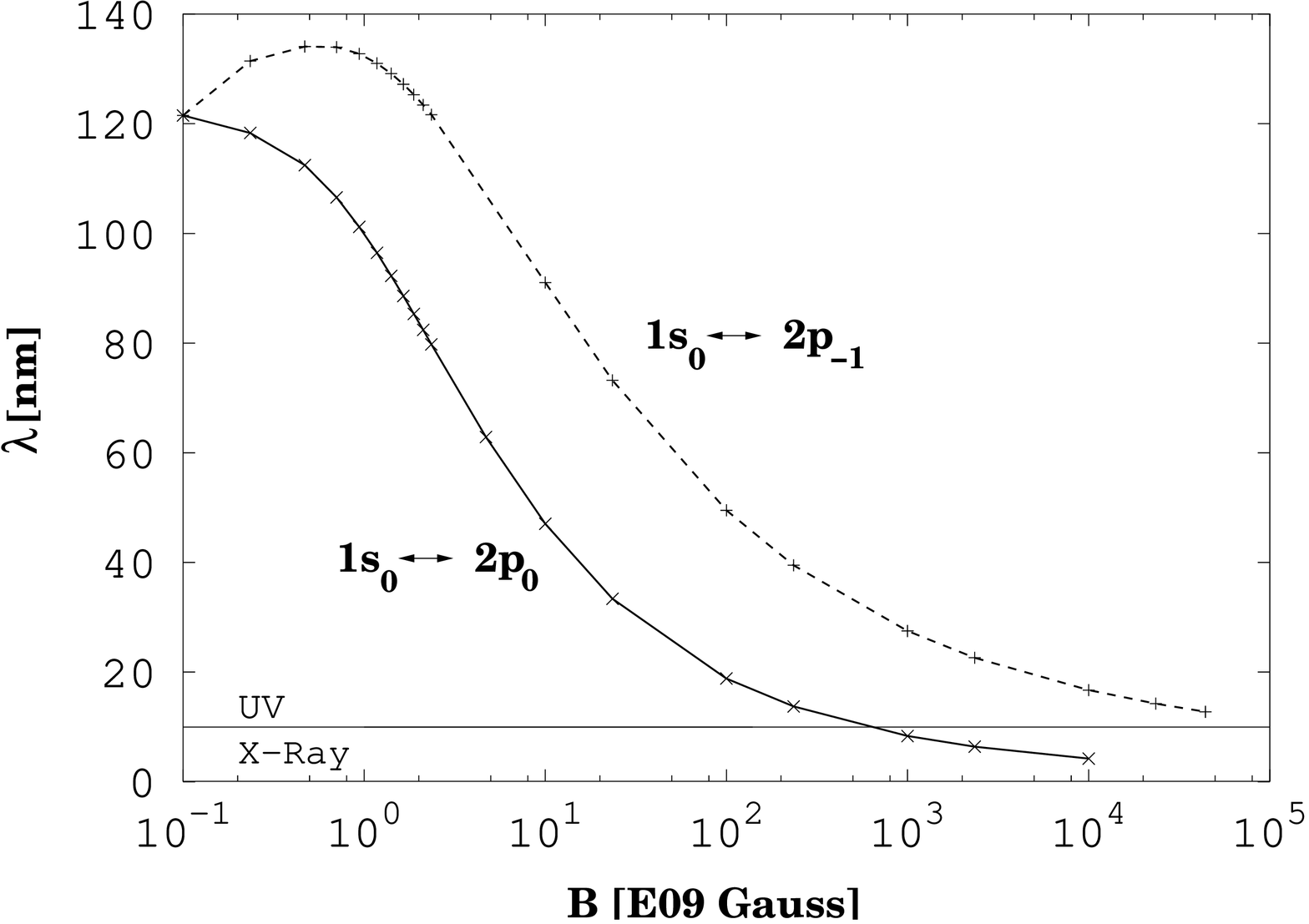}
       \caption{\label{fig:lambda}
Wavelengths of (a) the   right-circular  polarized radiation associated with the  transition  $1s_{0}   \leftrightarrow 2p_{-1}$ (dashed curve)
and (b) of the linearly polarized radiation associated with the transition $2p_{0} \leftrightarrow  1s_{0} $ (solid curve) as functions of the magnetic field $B$. The curves show the calculated values of  the wavelengths $\lambda_{1s_{0}   \leftrightarrow 2p_{-1}}$ (marked with the symbol $+$) and  $\lambda_{1s_{0}   \leftrightarrow 2p_{0}}$ (marked with the symbol $\times$)  joined by line-segments.}
\end{center}
\end{figure}

\subsection{Transitions}

With the approximate wavefunctions (\ref{trialgs}), (\ref{trial2pm1}) and (\ref{trial2p0})  found in the variational procedure described above, we carried out a study of the  electromagnetic transitions  between the states $2p_{-1}$, $2p_{0}$ and  the ground state
 $1s_{0}$, {\it i.e.}  $1s_{0}   \leftrightarrow 2p_{-1} \quad (\Delta m =1) $,
and  $ 1s_{0}
 \leftrightarrow 2p_{0} \quad (\Delta m =0)$.
The transition   $1s_{0}   \leftrightarrow 2p_{-1}$ occurs by absorption (emission) of  circular-right-polarized radiation ($q=+1$), while the transition    $ 1s_{0}  \leftrightarrow 2p_{0}$ occurs by absorption (emission) of linearly-polarized (along the magnetic field direction) radiation ($q=0$). Neither $2p_{-1}$, nor  $2p_{0}$ have a  excitation (de-excitation) mode to the ground state via  left-polarized radiation.

 The first remarkable observation concerning the electromagnetic transitions is the fact
 that,   if for small magnetic fields the transition probabilities  for both transitions
are comparable in magnitude, for strong magnetic
fields the (circularly polarized)  transition  $1s_{0}   \leftrightarrow 2p_{-1}$ is strongly suppressed in comparison to the corresponding
 transition probability for the (linearly polarized) transition $1s_{0}  \leftrightarrow 2p_{0}$ (see Fig.~(\ref{fig:w})). This
phenomenon is  a consequence of the strong deformation of the electronic distribution due to the enourmous Lorentz force acting on it,  being elongated in the direction of  the magnetic field, thus, enhancing the longitudinal polarization transition mode $q=0$ with respect to the transverse transition modes $q=\pm 1$.

In the other extreme of small magnetic fields, perturbative results for the  dipole and  oscillator strengths and transition probabilities  are given by \cite{Horaciothesis}:
\begin{eqnarray*}
d^{(+1)}_{1s_0\leftrightarrow  2p_{-1}} &\simeq&  0\textrm{.}555 + 29\textrm{.}776 B^2\,, \\
f^{(+1)}_{1s_0\leftrightarrow 2p_{-1}}  &\simeq& 0\textrm{.}416
                                - 0\textrm{.}555 B + 25\textrm{.}841B^2 \,,\\
              w^{(+1)}_{1s_0\leftrightarrow 2p_{-1}}&\simeq& (6\textrm{.}266
                                 - 25\textrm{.}066B+171\textrm{.}280 B^2)   \times10^{8}\sec^{-1}\,,
\end{eqnarray*}
for the   $1s_0\leftrightarrow  2p_{-1}$ transition, and
\begin{eqnarray*}
d^{(0)}_{1s_0\leftrightarrow  2p_{0}} &\simeq& 0\textrm{.}555 +9\textrm{.}388\,\,B^2\,,\\
f^{(0)}_{1s_0\leftrightarrow 2p_{0}}&\simeq& 0\textrm{.}416
                                           + 8\textrm{.}428B^2 \,,\\
  w^{(0)}_{1s_0\leftrightarrow 2p_{0}}&\simeq& (6\textrm{.}266
                                           + 126\textrm{.}894B^2)
               \times10^{8}\sec^{-1} \,,
\end{eqnarray*}
for the      $1s_0\leftrightarrow 2p_{0}$ transition, where the magnetic field strength $B$ is  in a.u.     These results show different behaviors for both transitions. While for the $1s_0\leftrightarrow  2p_{0}$ transition all quantities $d^{(0)}_{1s_0\leftrightarrow 2p_{0}},f^{(0)}_{1s_0\leftrightarrow 2p_{0}}$ and $w^{(0)}_{1s_0\leftrightarrow 2p_{0}}$ are growing functions of $B^2$, in the case of the  $1s_0\leftrightarrow  2p_{-1}$ transition the oscillator strength $f^{(+1)}_{1s_0\leftrightarrow  2p_{-1}}$ and the transition probability $w^{(+1)}_{1s_0\leftrightarrow  2p_{-1}}$ slightly decrease for  small increasing magnetic fields reaching a minimum for $B\simeq 10^{-2}$\,a.u.,  and $B\simeq 10^{-1}$\,a.u. respectively.
For larger magnetic fields $B\sim 10^{-1}$\,a.u. all quantities
 $d^{(0)}_{1s_0\leftrightarrow 2p_{0}},f^{(0)}_{1s_0\leftrightarrow 2p_{0}}, w^{(0)}_{1s_0\leftrightarrow 2p_{0}}$ and $d^{(+1)}_{1s_0\leftrightarrow  2p_{-1}},f^{(+1)}_{1s_0\leftrightarrow  2p_{-1}}, w^{(+1)}_{1s_0\leftrightarrow  2p_{-1}}$ are growing functions of $B^2$.
As we go to higher (non-perturbative) magnetic fields the results obtained with the variational functions (\ref{trialgs}), (\ref{trial2pm1}) and (\ref{trial2p0}) show that  the dipole strengths and the oscillator strengths of both transitions eventually reach a maximum as the magnetic field grows after which they start to decrease monotonously in the region of high magnetic fields $B\gtrsim 1$\,a.u.    For the  $1s_0\leftrightarrow 2p_{-1}$ transition
the maximum in the dipole strength $d^{(+1)}_{1s_0\leftrightarrow  2p_{-1}}(B)$ and the maximum in  the oscillator strength $f^{(+1)}_{1s_0\leftrightarrow  2p_{-1}}(B)$ approximately coincide at $B\simeq 0.3$\,a.u.   The transition probability also shows a maximum for the same value magnetic field. In contrast, in the  $1s_0\leftrightarrow 2p_{0}$ transition the maximum in the dipole strength $d^{(0)}_{1s_0\leftrightarrow 2p_{0}}(B)$ occurs for $B\simeq 0.3$\,a.u. but the maximum in the oscillator strength $f^{(0)}_{1s_0\leftrightarrow 2p_{0}}(B)$ occurs for $B\simeq 2$\,a.u. However, the transition probability $w^{(0)}_{1s_0\leftrightarrow 2p_{0}}(B)$  is an increasing function of the magnetic field (Fig.~(\ref{fig:w})). 								\\

Our results for $d^{(+1)},f^{(+1)}, w^{(+1)}$  corresponding to the $1s_{0}   \leftrightarrow 2p_{-1}$ transition  are in good agreement  with the results in ~\cite{Ruderbook} in the whole domain of magnetic fields. The relative differences between our results for the dipole and oscillator strengths and  the corresponding results in \cite{Ruderbook}  are $\sim 10^{-5}$   for magnetic fields $B\sim 0.1$\,a.u. increasing  rather monotonously when the magnetic field grows being $\sim 10^{-3}$ at $B\sim 1000$\,a.u. Our result for the transition probability are also in good agreement with the results in \cite{Ruderbook}, the relative differences  being $\sim 10^{-4}-10^{-3}$ for magnetic fields $B\sim 0.1- 100$\,a.u. The largest relative difference $\sim 10^{-2}$ is observed at $B=1000$\,a.u. (see Table~(\ref{Tabla:2P-1})).\\

Major differences occur in the case of the $1s_{0}   \leftrightarrow 2p_{0}$ transition.
For instance, the relative differences between present results for the oscillator strengths and  the corresponding results in \cite{Ruderbook}  are   $\sim 10^{-4}$   for magnetic fields $B\sim 0.1$\,a.u. increasing with a magnetic field increase being
  $\sim 10^{-1}$  at $B\sim 1000$\,a.u.  The corresponding relative differences for the dipole strength are $\sim 10^{-4}$
for magnetic fields $B\sim 0.1$\,a.u. increasing up to $\sim 0.3$ at $B=1000$\,a.u. A similar deviation  is observed in the results for the    transition probability for  $1s_{0}   \leftrightarrow 2p_{0}$   where the relative difference is $\sim 10^{-4}$
for magnetic fields $B\sim 0.1$\,a.u. and increasing up to $\sim 0.3$ at $B=1000$\,a.u.
It is important to  emphasize here that there exist no criterion for other observables, but the case of the total (or binding) energy,  to decide about which values are more accurate.
\\

In the domain of very strong magnetic fields ($B\gtrsim 10^{12}$\,G) our results for   dipole strengths of both transitions show the approximate  scalings
$d^{(0)}_{1s_0\leftrightarrow  2p_{0}}  \sim \left(\frac{B}{B_0}\right)^{-2/3}$, and $d^{(+1)}_{1s_0\leftrightarrow  2p_{-1}}  \sim  \left(\frac{B}{B_0}\right)^{-1}$,  accurate in  $\lesssim 10\%$ and $\lesssim 1\%$ respectively.
Likewise, the results for the oscillator strengths scale approximately as $f^{(0)}_{1s_0\leftrightarrow  2p_{0}}  \sim \frac{3}{2}\left(\frac{B}{B_0}\right)^{-1/3}$, and $f^{(+1)}_{1s_0\leftrightarrow  2p_{-1}}  \sim   \left(\frac{B}{B_0}\right)^{-4/5}$,  both accurate in  $\lesssim 5\%$.\\

Finally, in Fig.~(\ref{fig:lambda}) we show the wavelengths $\lambda = {2\pi}/{\alpha\,(E^{\tau'}_{b} - E^{\tau}_{b})}$ ($\alpha\simeq 1/137$ is the fine structure constant and $(E^{\tau'}_{b} - E^{\tau}_{b})$  is the (variational) energy difference between the initial and final states) of the electromagnetic radiation associated with each transition as function of the magnetic field strength $B$.   From Fig.~(\ref{fig:lambda}), we can see that the wavelength of the longitudinal polarized radiation in the transition   $1s_{0} \leftrightarrow 2p_{0}$ is a monotonously decreasing function of the magnetic field, reaching the domain of X-rays for $B\sim 10^{12}$\,G, while    the wavelength of the right polarized radiation corresponding to the transition   $1s_{0} \leftrightarrow 2p_{-1}$ increases for small to weak magnetic fields reaching a maximum\footnote{Hydrogen transitions whose wavelengths go through maxima or minima as functions of the magnetic field are called {\it stationary lines}.} for  $B\sim 0.2$\,a.u. (with $\lambda_{max}\sim 1340.7$~\AA), and decreasing for larger magnetic fields. Yet, this transition remains always visible in the UV-region even for the stronger magnetic fields considered in the present study $B\sim 10000$\,a.u.

Figures (\ref{fig:DS}), (\ref{fig:OS}) and  (\ref{fig:w}) show  the results of the calculations for the dipole strengths, oscillator strengths and transition probabilities  (formulas (\ref{dtautau}), (\ref{ftautau}) and (\ref{wtautau}) respectively)   for the  transitions  $1s_{0}   \leftrightarrow 2p_{-1}$, and  $ 1s_{0}  \leftrightarrow 2p_{0}$ in the domain of magnetic fields $B\sim 0.1 -10000$\,a.u.  Tables (\ref{transitions2P1-1S0}) and  (\ref{transitions2P0-1S0}) also show the results of those quantities for magnetic fields $B=0-10000$\,a.u.

\section{Conclusions}

Along the present study we have used a variational approach with a physics recipe for choosing simple trial functions, as a test for an alternative method to
study  electromagnetic transitions in the hydrogen atom placed in a constant magnetic
field. We assume that the proton is infinitely massive (Born-Oppenheimer approximation of zero order). It was found that the method yields very accurate results for the binding energies, in particular, for the lowest states studied $1s_0, 2p_{-1},2p_0$, in all the range of  magnetic fields $B=0-4.414\times 10^{13}$\,G. The results for binding energies  show that the accuracy given by  the simple 7-parametric trial functions (\ref{trialgs}),(\ref{trial2pm1}) and (\ref{trial2p0}) is excellent for {\em small} magnetic fields $B\lesssim 0.1$\,a.u., however it  decreases monotonously as the magnetic field grows: the relative differences between the results given by the proposed trial functions (\ref{trialgs}),(\ref{trial2pm1}) and (\ref{trial2p0}) and the today's most accurate results \cite{Ruderbook} and \cite{K&L:ET} reach about $\sim 1\%$ for $B=1000\,$a.u.  A possible explanation for  this reduction in  accuracy is the  fact that transverse and longitudinal degrees of freedom ($\rho$ and $z$ respectively)  in  the trial functions (\ref{trialgs}), (\ref{trial2pm1}) and (\ref{trial2p0}) appear {\lq isotropically\rq} in the electron-proton distance $r$, preventing their adiabatic separation at $B\to \infty$.

Dipole $d^{(q)}(B)$, and oscillator strengths $f^{(q)}(B)$, of the   electromagnetic transitions ($1s_{0}   \leftrightarrow 2p_{-1}$ and  $ 1s_{0}  \leftrightarrow 2p_{0}$)  were computed  with the approximate  wave functions (\ref{trialgs}),(\ref{trial2pm1}), (\ref{trial2p0}) as functions of the magnetic field strength $B$. Our results for the $1s_{0}   \leftrightarrow 2p_{-1}$ transition are in very good agreement  with the results of ~\cite{Ruderbook}, with small deviations varying rather monotonously in $0.001-0.1\%$ for the interval of magnetic fields $B=0.1-1000$\,a.u.    Major deviations between the present results and the results in~\cite{Ruderbook}  for the dipole and oscillator strengths were observed in the case of the $1s_{0}   \leftrightarrow 2p_{0}$  transition, where we have differences up to $\sim 30\%$ in the oscillator strength $f^{(0)}_{1s_{0}   \leftrightarrow 2p_{0}} (B)$ at $B=1000$\,a.u. A similar difference $\sim 30\%$ is obtained for the corresponding  transition probability  $w^{(0)}_{1s_{0}   \leftrightarrow 2p_{0}} (B)$. It is worth to emphasize that for a strong magnetic field $B\simeq 1000$\,a.u. the   transition probability $w^{(0)}_{1s_{0}   \leftrightarrow 2p_{0}} (B)$ is 3 orders of magnitude larger than the transition probability  $w^{(+1)}_{1s_{0}   \leftrightarrow 2p_{-1}} (B)$   corresponding to the $1s_{0}   \leftrightarrow 2p_{-1}$ transition, so  the difference in results for the transition probability  $w^{(+1)}_{1s_{0}   \leftrightarrow 2p_{-1}} (B)$ might be of relevance for the analysis of the X-ray spectra of neutron stars.

It is important to mention that if for the variational binding energies there exists a criterion to decide which results given by different approximate wavefunctions are better (since variational binding energies approach from below to the true binding energies),  there exists no similar criterion for other expectation values or matrix elements. Thus, it is not clear so far which  results for dipole strengths  are better. Therefore, more investigations on the electromagnetic transitions in the hydrogen atom in a magnetic field, specially in the domain of strong magnetic fields,  would be desirable in order to answer this question.

\begin{acknowledgments}
  This work was supported in part by DGAPA grant
  {\scshape PAPIIT {\bf IN121106}} (Mexico). The authors are heartily grateful to
A. Turbiner for his valuable and numerous discussions, and for his careful readings of the manuscript.
\end{acknowledgments}

\begin{widetext}

\begingroup
\squeezetable
\begin{table*}
\begin{tabular}{ r@{.}l | r@{.}l || r@{.}l |l |l }
\hline
\hline
\multicolumn{2}{l}{}\vline
  &\multicolumn{4}{c}{Variational calculation} \vline
  &\multicolumn{1}{c}{Ref.~\cite{Ruderbook}} \vline
  &\multicolumn{1}{c}{Ref.~\cite{K&L:ET}}\\
\multicolumn{2}{l}{}\vline
  &\multicolumn{4}{l}{} \vline
  &\multicolumn{1}{l}{}\vline
  &\multicolumn{1}{c}{} \\
\hline
\multicolumn{2}{l}{$B\times10^{9}G$} \vline
  &\multicolumn{2}{c}{$E_T$ (Ry)} \vline\,\vline
  &\multicolumn{4}{c}{$E_b$ (Ry)}\\
\hline
\hline
0&235   &    -0&99505296 &  1&09505296\footnote[1]{Results of Ref.~\cite{PotekhinTurbiner}.}\hspace{0.4cm} & 1.095053 & 1.09505296\\
1&0     &    -0&9208225  &  1&3463544 \footnotemark[1] & -        & -          \\
2&35    &    -0&662332   &  1&662332\footnotemark[1] & 1.662338 & 1.66233779 \\
10&0    &     1&640362   &  2&614957 \footnotemark[1]{}  & -        & -          \\
23&5    &     6&50522    &  3&4948   \footnotemark[1]{}  & 3.495594 & 3.49559433 \\
100&0   &    36&8398     &  5&7134  \footnotemark[1]{}   & -        & -          \\
235&0   &    92&4356     &  7&564   \footnotemark[1]{}   & 7.5781   & 7.57960847 \\
1000&0  &   413&662      & 11&870    \footnotemark[1]{}   & -        & -          \\
2350&0  &   984&773      & 15&23     \footnotemark[1]{}   &15.3241   &15.32484649 \\
10000&0 &  4232&77       & 22&55   \footnotemark[1]{}    & -        & -          \\
23500&0 &  9972&0        & 27&96                         & -        & -          \\
44140&0 & 18750&5        & 32&5  \footnotemark[1]{}      & -        & -          \\
\hline
\hline
 \end{tabular}
\caption{Total $E_T$ and binding $E_b$ energies for the ground state $1s_0$ of the hydrogen atom in a magnetic field calculated with  the variational function (\ref{trialgs}) compared with the results obtained by Ruder {\it et al.}~\cite{Ruderbook} and Kravchenko {\it et al.}~\cite{K&L:ET}. The values of the energies have been rounded to the first two non-coinciding digits respect to the values of \cite{K&L:ET}.}
 \label{Tabla:1S0}
\end{table*}
\endgroup
\end{widetext}


\begin{widetext}
\begingroup
\squeezetable
\begin{center}
\begin{table*}
\begin{center}
\begin{tabular}{ r@{.}l | r@{.}l || r@{.}l | l | l }
\hline
\hline
\multicolumn{2}{l}{}\vline
  &\multicolumn{4}{c}{Variational Calculations} \vline
  &\multicolumn{1}{c}{Ref.~\cite{Ruderbook}} \vline
  &\multicolumn{1}{c}{Ref.~\cite{K&L:ET} }\\
\multicolumn{2}{l}{}\vline
  &\multicolumn{4}{l}{} \vline
  &\multicolumn{1}{l}{} \vline
  &\multicolumn{1}{c}{} \\
\hline
\multicolumn{2}{l}{$B\times10^{9}G$} \vline
  &\multicolumn{2}{c}{$E_T$ (Ry)} \vline\,\vline
  &\multicolumn{4}{c}{$E_b$ (Ry)}\\
\hline
\hline
0&235   &    -0&3016912  &  0&40169120 \hspace{0.4cm} &  0.4016913 &  0.40169135\\
1&0     &    -0&232205   &  0&657737   &   -        &  -           \\
2&35    &     0&08684    &  0&913163   &  0.9131941 &  0.91319412  \\
10&0    &     2&64128    &  1&6140     &   -        &  -           \\
23&5    &     7&75028    &  2&2497     &  2.250845  &  2.25084468  \\
100&0   &    38&6802     &  3&873      &   -        &  -           \\
235&0   &    94&744      &  5&256      &  5.26948   &  5.26952133  \\
1000&0  &   416&975      &  8&557      &   -        &  -           \\
2350&0  &   988&80       & 11&20       & 11.27681   & 11.27684216  \\
10000&0 &  4238&22       & 17&10       &   -        &  -           \\
23500&0 &  9978&44       & 21&56       &   -        &  -           \\
44140&0 & 18757&6        & 25&4        &   -        &  -           \\
\hline
\hline
\end{tabular}
\end{center}
\caption{Total $E_T$ and binding $E_b$ energies for the state $2p_{-1}$ of the hydrogen atom in a magnetic field
as given by the trial function (\ref{trial2pm1}) compared with the results obtained by Ruder {\it et al.}~\cite{Ruderbook} and Kravchenko {\it et al.}~\cite{K&L:ET}. The values of the energies have been rounded to the first two non-coinciding digits respect to the values of \cite{K&L:ET}.}
\label{Tabla:2P-1}
\end{table*}
\end{center}
\endgroup
\end{widetext}


\begin{widetext}

\begingroup
\squeezetable
\begin{center}
\begin{table*}
\begin{center}
\begin{tabular}{ r@{.}l | r@{.}l || r@{.}l | l  | l }
\hline
\hline
\multicolumn{2}{l}{}\vline
  &\multicolumn{4}{c}{Variational Calculations} \vline
  &\multicolumn{1}{c}{Ref.~\cite{Ruderbook}} \vline
  &\multicolumn{1}{c}{Ref.~\cite{K&L:ET}}\\
\multicolumn{2}{l}{}\vline
  &\multicolumn{4}{l}{}  \vline
  &\multicolumn{1}{l}{} \vline
  &\multicolumn{1}{c}{} \\
\hline
\multicolumn{2}{l}{$B\times10^{9}G$} \vline
  &\multicolumn{2}{c}{$E_T$ (Ry)} \vline\hspace{-0.09cm} \vline
  &\multicolumn{4}{c}{$E_b$ (Ry)}  \\
\hline
\hline
0&235   &   -0&2248199  & 0&3248199 \hspace{0.4cm} & 0.3248202 & 0.32482016 \\
1&0     &   -0&008703   & 0&434235   & -         & -          \\
2&35    &    0&48008    & 0&5199     & 0.5200132 & 0.52001323 \\
10&0    &    3&5779     & 0&6774     & -         & -          \\
23&5    &    9&2359     & 0&7641     & 0.7652975 & 0.76529970 \\
100&0   &   41&674      & 0&8796     & -         & -          \\
235&0   &   99&074      & 0&9255     & 0.9272354 & 0.92723552 \\
1000&0  &  424&561      & 0&9710     & -         & -          \\
2350&0  &  999&016      & 0&9844     & 0.9849900 & 0.9849900  \\
10000&0 & 4254&33       & 0&9938     & -         & -          \\
\hline
\hline
\end{tabular}
\end{center}
\caption{Total $E_T$ and binding $E_b$ energies for the state $2p_{0}$ of the hydrogen atom in a magnetic field
as given by the trial function (\ref{trial2p0}) compared with the results obtained by Ruder {\it et al.}~\cite{Ruderbook} and Kravchenko {\it et al.}~\cite{K&L:ET}. The values of the energies have been rounded to the first two non-coinciding digits respect to the values of \cite{K&L:ET}.}
\label{Tabla:2P0}
\end{table*}
\end{center}
\endgroup
\end{widetext}


 \begingroup
\squeezetable
\begin{table*}
\begin{center}
\begin{tabular}{l|ll|ll|ll}
\hline
\hline
        &\multicolumn{2}{c}{Dipole Strength} \vline
        &\multicolumn{2}{c}{Oscillator Strength} \vline
        &\multicolumn{2}{c}{Transition}   \\
        &\multicolumn{2}{c}{$d^{(+1)}_{1s_{0}\leftrightarrow 2p_{-1}}$} \vline
        &\multicolumn{2}{c}{$f^{(+1)}_{1s_{0}\leftrightarrow 2p_{-1}}$} \vline
&\multicolumn{2}{c}{Probability $w^{(+1)}_{1s_{0}\leftrightarrow 2p_{-1}}$} \\
\hline
     $B\times10^{9}G$& & Ref.~\cite{Ruderbook} & & Ref.~\cite{Ruderbook} & & Ref.~\cite{Ruderbook} \\
\hline
\hline
0.0     &0.55493     &            & 0.41620    &            & 6.2664 &    \\
0.235   &0.64837     & 0.6484     & 0.44955    &  0.4496    & 5.7849 & 5.7852 \\
1.0     &0.68681     & -          & 0.47295    &  -         & 6.0030 & -  \\
2.35    &0.50133     & 0.5015     & 0.37558    &  0.3757    & 5.6423 & 5.6437 \\
10.0    &0.18073     & -          & 0.18089    &  -         & 4.8509 & -  \\
23.5    &0.08576     & 0.08584    & 0.10678    &  0.1069    & 4.4305 & 4.4313 \\
100.0   &0.021987    & -          & 0.04046    &  -         & 3.6683 & -  \\
235.0   &0.009581    & 0.009591   & 0.02212    &  0.02214   & 3.1552 & 3.1587 \\
1000.0  &0.002296    & -          & 0.00761    &  -         & 2.2355 & -  \\
2350.0  &0.0009837   & 0.0009847  & 0.00396    &  0.003985  & 1.7231 & 1.7474 \\
10000.0 &0.0002325   & -          & 0.00127    &  -         & 1.0068 & -  \\
23500.0 &0.0000992   & -          & 0.00063    &  -         & 0.6950 & -  \\
44140.0 &0.0000529   & -          & 0.00038    &  -         & 0.5165 & -  \\
\hline
\hline
\end{tabular}
\caption{\label{transitions2P1-1S0} Results for the
  electromagnetic  transition  $ 1s_0\leftrightarrow 2p_{-1}$ in the hydrogen atom in a magnetic field $B$ obtained with the variational functions (\ref{trialgs}) and (\ref{trial2pm1}) compared with the results of Ruder {\it et al.} \cite{Ruderbook} for the case of infinite nuclear mass.   The units for the transition probability  are $10^{8}\sec^{-1}$.}
\end{center}
\end{table*}
\endgroup

\begingroup
\squeezetable
\begin{table}[h]
\begin{tabular}{l|ll|ll|ll}
\hline
\hline
        &\multicolumn{2}{c}{Dipole Strength} \vline
        &\multicolumn{2}{c}{Oscillator Strength} \vline
        &\multicolumn{2}{c}{Transition } \\
        &\multicolumn{2}{c}{$d^{(0)}_{1s_{0}\leftrightarrow 2p_{0}} $} \vline
        &\multicolumn{2}{c}{$f^{(0)}_{1s_{0}\leftrightarrow 2p_{0}} $} \vline
 &\multicolumn{2}{c}{Probability $w^{(0)}_{1s_{0}\leftrightarrow 2p_{0}} $}\\
\hline
     $B\times10^{9}G$& & Ref. \cite{Ruderbook} & & Ref. \cite{Ruderbook} & & Ref. \cite{Ruderbook} \\
\hline
\hline
0.0     &0.55493   &         &  0.41620  &           &  6.2664    &          \\
0.235   &0.60843   & 0.6083  &  0.46864  &   0.4685  &  7.4417    & 7.4401 \\
1.0     &0.67290   & -       &  0.61377  &  -        &  13.668   & -        \\
2.35    &0.58565   & 0.5902  &  0.66905  &   0.6742  &  23.372   & 23.549 \\
10.0    &0.33129   & -       &  0.64189  &  -        &  64.499   & -        \\
23.5    &0.21101   & 0.2252  &  0.57621  &   0.6149  &  115.01   & 122.69\\
100.0   &0.088027  & -       &  0.42551  &  -        &  266.12  & -        \\
235.0   &0.050464  & 0.06217 &  0.33502  &   0.4135  &  395.24  & 489.56\\
1000.0  &0.019169  & -       &  0.20893  &  -        &  664.29  & -        \\
2350.0  &0.010898  & 0.01585 &  0.15522  &   0.2273  &  842.82  & 1250.81\\
10000.0 &0.0043462 & -       &  0.093697 &  -        &  1165.6 & -        \\
\hline
\hline
\end{tabular}
\caption{\label{transitions2P0-1S0} Results for the
  electromagnetic  transition  $1s_0\leftrightarrow 2p_{0} $ in the hydrogen atom in a magnetic field $B$ obtained with the variational functions (\ref{trialgs}) and (\ref{trial2p0}) compared with the results of Ruder {\it et al.} \cite{Ruderbook} for the case of infinite nuclear mass.
  The units for the transition probability  are $10^{8}\sec^{-1}$.}
\end{table}
  \endgroup

\end{document}